\documentclass[ a4paper, traditabstract]{aa}

\usepackage{graphicx}
\usepackage[usenames]{color}

\usepackage[breaklinks, colorlinks, citecolor=blue]{hyperref}

\usepackage[nonamebreak]{natbib}
\usepackage{longtable,lscape}
\usepackage{txfonts} 
\usepackage{amsfonts}
\usepackage{amssymb}
\usepackage{newlfont}
\usepackage{textcomp}
\usepackage{times}
\usepackage{units}
\usepackage{paralist}
\usepackage{txfonts}
\usepackage{sidecap}

\usepackage{fixltx2e}

\def\reff@jnl#1{{\rm#1\/}}
\def\apj{\reff@jnl{ApJ}}       
\def\apjs{\reff@jnl{ApJS}}     
\def\aaps{\reff@jnl{A\&AS}}    
\def\mnras{\reff@jnl{MNRAS}}   
\def\prd{\reff@jnl{Phys.\ Rev.\ D}}    

\newcommand{\beq}{\begin{equation}}
\newcommand{\eeq}{\end{equation}}

\newcommand{\be}{\begin{equation}}
\newcommand{\ee}{\end{equation}}
\newcommand{\bea}{\begin{eq}}
\newcommand{\eea}{\end{equation}}

{\begin{enumerate}\setlength{\itemsep}{0mm}}%
{\end{enumerate}}
{\begin{enumerate}\setlength{\itemsep}{0mm}}%
{\end{enumerate}}

\newcommand{\bc}{\begin{center}}
\newcommand{\ec}{\end{center}}
\newcommand{\bi}{\begin{itemize}}
\newcommand{\ei}{\end{itemize}}
\newcommand{\ben}{\begin{enumerate}}
\newcommand{\een}{\end{enumerate}}

\usepackage{mathrsfs}
\usepackage{mathbbol}

\newfont{\gwpfont}{cmssq8 scaled 1000}

\def\lesssim{\mathrel{\hbox{\rlap{\hbox{\lower4pt\hbox{$\sim$}}}\hbox{$<$}}}}
\def\gtrsim{\mathrel{\hbox{\rlap{\hbox{\lower4pt\hbox{$\sim$}}}\hbox{$>$}}}}

\begin{document}

   \title{A new parameterization of the reionisation history}

   \abstract{Motivated by the current constraints on the epoch of
     reionisation from recent cosmic microwave background
     observations, ionising background measurements of star-forming
     galaxies, and low redshifts line-of-sight probes, 
     we propose a new{ data-motivated} parameterisation of
     the history of the average ionisation fraction. This
     parameterisation describes a flexible redshift-asymmetric
     reionisation process in two regimes that is capable of fitting all the current
     constraints.}


\authorrunning{Douspis et al.}
\titlerunning{A new parameterization of the reionisation history}

\author{\small M. Douspis\inst{1} and N. Aghanim\inst{1} and S. Ili\'c\inst{1,2} and  M. Langer\inst{1}}
\institute{$^1$Institut d'Astrophysique Spatiale, Universit\'e Paris-Sud, CNRS , UMR8617, Orsay, F-91405, France\\ $^2$IRAP, Universit\'e de Toulouse, UPS-OMP, CNRS, 14, avenue Edouard Belin, F-31400 Toulouse, France}

   \maketitle



\section{Introduction}
After recombination at $z\approx1090$, the universe was essentially
neutral. Observations of the Gunn-Peterson (GP) effect
\citep{Gunn65} in quasar spectra \citep[e.g.][]{Fan06} indicate
that intergalactic gas has become almost fully reionised by redshift
$z\sim6$. This transition, called cosmic reionisation, is one of the
most important events in cosmic structure formation and in cosmology. It is also
of particular importance since it may be related to many fundamental
questions, such as the presence of annihilating or decaying dark
matter particles, properties of the first galaxies, physics of
(mini-)quasars, and the formation of very metal-poor stars. The reionisation
process starts with the formation of the first generation of early
star-forming galaxies and quasars that emit ultraviolet radiation
that reionises the neutral regions around them. This is the so-called
patchy reionisation stage \citep{agh96}. After a sufficient number of
ionising sources have formed and ionised regions have overlapped, the
ionised portion of the gas in the Universe rapidly increases until
hydrogen becomes fully ionised. This period, during which the cosmic
gas went from neutral to ionised, is known as the Epoch of
Reionisation (EoR).  The transition from neutral to ionised diffuse
gas is still not observed directly but is constrained by the absorption spectra of
very distant quasars and gamma ray bursts, which reveal neutral hydrogen
in intergalactic clouds. These  GP effect observations show that the diffuse
gas in the Universe is mostly ionised up to a redshift of order 6
\citep{Fan06}.

The reionisation process describes the balance between the
recombination of free electrons with protons to form neutral hydrogen
and the ionisation of hydrogen atoms by photons with energies $h\nu >
13.6$~eV.  The first empirical, analytic, and numerical models of the
reionisation process \citep[e.g.][]{agh96,gru98,mad99,gne00,cia03}
highlighted the basic physics that give rise to the ionised intergalactic
medium (IGM) at late times and provided predictions of the effects on the
cosmic microwave background (CMB). The reionisation history is conveniently expressed
in terms of the filling factor of ionised hydrogen $Q_\mathrm{HII}$.
One of the relevant physical and the most
commonly used quantities for characterising reionisation is the Thomson
scattering optical depth to the CMB $\tau = \int_{\eta_\mathrm{CMB}}^{\eta_0} a n_e
\sigma_\mathrm{T}\,\mathrm{d}\eta$, where $n_e$ is the number density
of free electrons at a conformal time $\eta$, $\sigma_\mathrm{T}$ is
the Thomson scattering cross-section, $a$  the scale factor, and
$\eta_0$  the conformal time today. 

Reionisation leaves imprints on the CMB
power spectra, both in polarisation and in intensity, and through the
kinetic Sunyaev-Zeldovich (kSZ) effect, due to the re-scattering of
photons off free electrons \citep[see][and references therein]{agh08}.
The temperature power spectrum is damped at scales larger than the
horizon. This effect is degenerate with the amplitude
$A_s$ and slope $n_s$ of the initial power spectrum. The CMB
temperature thus constrains the amplitude $A_se^{-2\tau}$. The
large-scale polarisation spectra present a reionisation ``bump'' that
WMAP first measured in TE in 2003 \citep{WMAP1}. In fact, CMB
observations allow us to shed light only on the  Thomson
optical depth $\tau$. In addition to the CMB, other astrophysical probes
constrain reionisation. 

Direct measurements of the ionisation state of the Universe through
Ly$\alpha$ emission or quasars, between redshift 5 and redshift 10,
have improved in quality and statistics
\citep[e.g.][]{Fan06,Faisst14}. Ionising background estimates from
star-forming galaxies improved as well in parallel to the depth of
observations \citep[e.g.][hereafter ISHI15, ROB15, BOU15,
respectively]{Ishi15,Rob15,Bou15}. New probes are becoming available
and increasingly useful, such as the 21cm brightness fluctuations
\citep[e.g.][]{Bowman10} or gamma-ray bursts
\citep[e.g.][]{Chornock14}. While it was hard to reconcile the
ionising power of galaxies with the high value of $\tau$ derived from
WMAP \citep{WMAP1}, the latest results from Planck \citep[][hereafter
PCP15]{PCP15} give an almost coherent picture of the EoR where
star-forming galaxies play a dominant role in ionising the Universe.

Large numbers of detailed models of the reionisation history have been
constructed to describe the transition from a neutral to an ionised
state and to reproduce the EoR observables
\citep{PLW10,Pan11,Mitra11,Mitra12,Mitra15}. These physically
motivated models need to make assumptions on various complex processes
(e.g. radiative transfer, star formation and evolution, chemical
enrichement) captured in effective parameters, such as escape
fraction, clumping factor, and star formation efficiency. As succesful
as they are, those models cannot be used easily to explore the global
history of the reionisation transition rapidly and without any prior
on the detailed physics.

In this letter, therefore, we provide a new parameterisation of the
mean ionised fraction of the universe as a function of redshift. We
exploit in particular the recent results based on the measurements of
UV and IR flux of early galaxies in order to derive a simple
data-motivated, fast, and economic expression of the ionised fraction.
It agrees with the current observational constraints on EoR better
than the standard parametisation used in CMB analyses.


\section{Observational status and current constraints on reionisation}

The reionisation process is intimately linked to the formation and
evolution of the first emitting sources. At redshifts as high as 15 to 30,
low-mass halos of matter detach from the general expansion, collapse,
and virialise. If they reach  high enough temperatures to excite
the lowest electronic levels of hydrogen, they cool and eventually form
Population III stars and dwarf atomic-cooling galaxies.
These emitting
sources,  thought to have initiated the reionisation process,
have different various properties (masses, soft/hard spectra, UV
escape fractions, etc.). The reionisation process is thus quite
complex since it implies multiple sources of UV photons that
combine their relative efficiencies at different epochs and
scales.

\begin{figure}[!th]
        \centering
        \includegraphics[width=8cm]{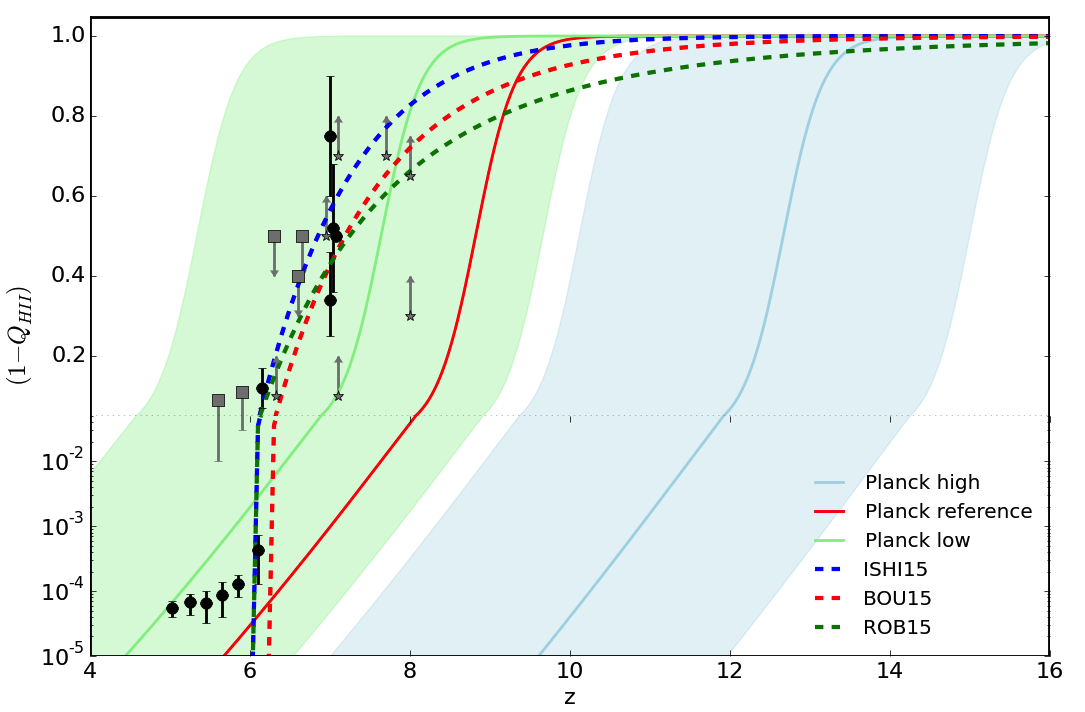}\\
\vspace{-0.0cm}\includegraphics[width=8cm]{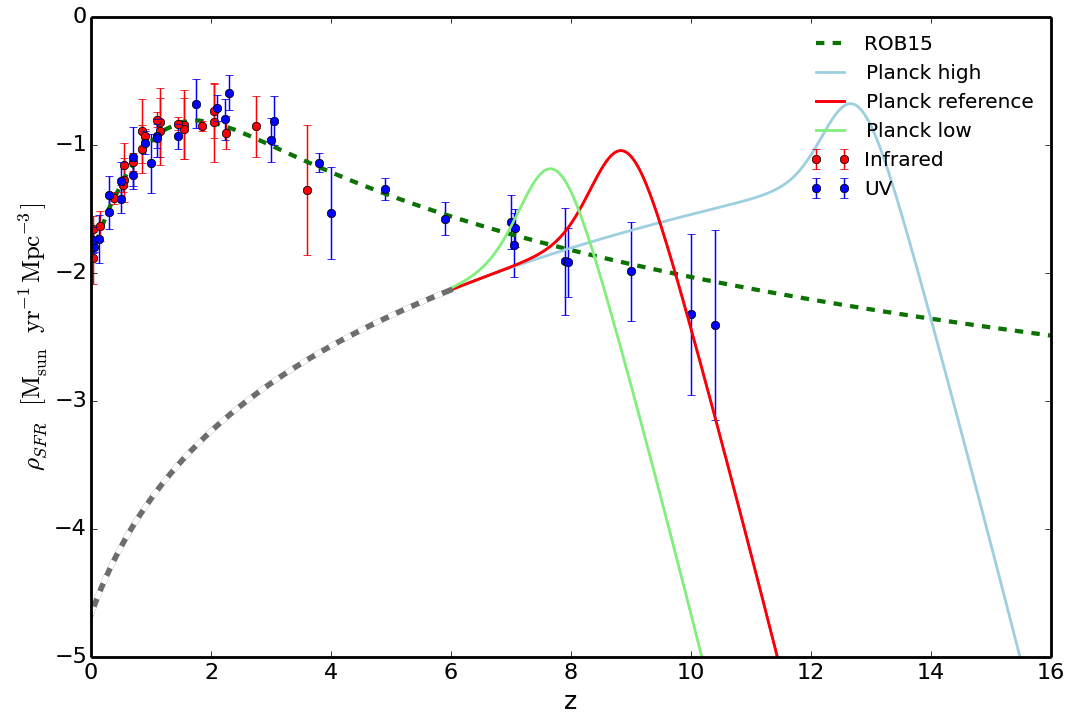}\\
\vspace{-0.1cm}\includegraphics[width=8cm]{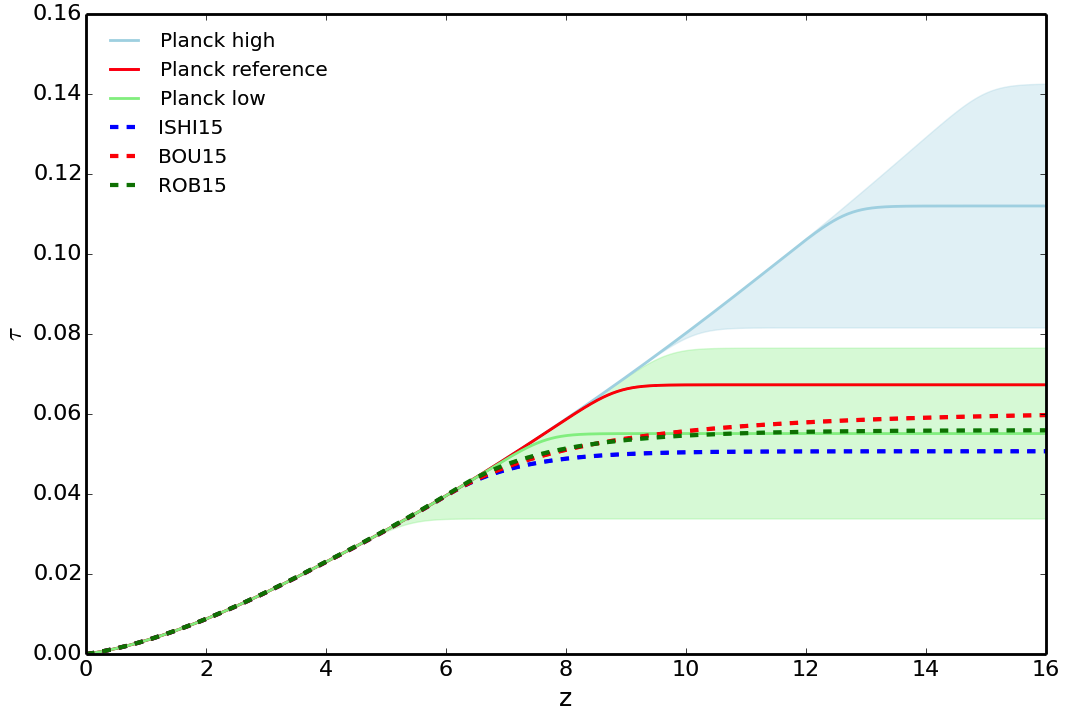}
\caption{Upper panel: Reionisation histories compared to low-redshift
  probes. Shaded envelopes show the low/high-$\tau$ values
  derived from Planck with uncertainties. The red line indicates the best
  Planck value $\tau =0.66$. Dashed lines are derived from the best
  ionisation rate ($\dot{N}_\mathrm{ion}$) of ISHI15, ROB15, BOU15. Symbols are
  current constraints from direct line-of-sight measurements compiled
  by BOU15. Middle panel: Corresponding $\rho_\mathrm{SFR}$ for Planck and the
  best fit of ROB15 with observational data compiled by ROB15. The
  grey dashed part corresponds to $Q_\mathrm{HII} \sim 1$. Lower panel: Evolution
  of the integrated optical depth for the different studies plotted in the
  upper panel.}
        \label{fig:all}
\end{figure}

The abundance of quasars declines rapidly beyond redshift
$z\sim6$. Quasars therefore cannot be significant contributors to
reionisation at high redshifts \citep[e.g.][]{Will10,Font12}, but they
are key players at lower redshifts since they ensure the second
reionisation of Helium \citep{McQ09}.  Faint  star-forming galaxies at
redshifts $z\gtrsim6$ have therefore been postulated as the most
likely sources of reionisation, and their time-dependent abundance and
spectral properties are crucial ingredients for understanding how
intergalactic hydrogen became reionised \citep[for reviews,
  see][]{Fan06,Rob10,Loeb13}. The high-redshift galaxy UV luminosity
function also provides a competitive observational constraint on
EoR. Based on the 2012 Hubble Ultra Deep Field
galaxies at $z\sim7-12$, \citet{Ellis13} show a continuous decline in
the abundance of star-forming galaxies over $6<z<10,$ whereas
\citet{Dun13}, based on the constancy of the UV continuum slope
measured in $z\simeq7-9$ galaxies over a wide range in luminosity,
show that stars at these redshifts are already enriched by earlier
generations, supporting the idea of a reionisation process that
extended beyond $z=9$. Finally, \citet{Rob13} combined the results
from UDF12 with the measured Thomson optical depth and stellar mass
density measurements \citep{Stark13} and provided constraints on the
role of high-redshift star-forming galaxies in the reionisation
process. Under reasonable assumptions about the escape fraction of
ionising photons and the IGM clumping factor, they find that the
observed galaxy population ($M_{UV}<-17$ at $z\sim8$) cannot
simultaneously reionise the universe by $z\sim6$ and produce a large
Thomson optical depth, unless the abundance of star-forming galaxies
and/or the ionising photon escape fraction increase in the range
$z\sim15-25$ with respect to what is currently observed \citep[see
  e.g.][]{kuh12}. The star formation rate (SFR) history allows the duration of reionisation to
be probed \citep{MD14} because star-forming
galaxies provide most of the ionising photons \citep{Rob13}. Recent works
by ISHI15, ROB15, and BOU15 (see also references therein) have shown
that star-forming galaxies at $z\sim 8-10$ were abundant
enough to ionise the Universe and yield a Thomson optical depth of the order of
0.06.

CMB data can also constrain reionisation history through secondary
anisotropies in temperature and polarisation. Measuring the latter has
proven to be a challenging task because the signal is strongly
contaminated by both foregrounds and systematics on large angular
scales \citep{PolP15}. This has affected the accuracy of optical depth
measurements \citep{WMAP1, WMAP3} and limited its precision
\citep{WMAP9,PCP15}. Recent results from Planck combining temperature
and polarisation provide $\tau=0.078\pm0.019$ in the $\Lambda$-CDM
model (PCP15). The degeneracy between $A_s$ and $\tau$ can be broken
by adding information about lensing \citep{PCP13}. Planck also provides
$\tau=0.66\pm0.013$ when considering temperature, polarisation,
lensing, and  baryonic acoustic oscillations (BAO) in $\Lambda$-CDM (Eq. 17e of PCP15, hereafter referred to as the Planck
reference). For an instantaneous reionisation, this
corresponds to a redshift of $z_{re}=8.8^{1.1}_{-1.2}$.  Extending the
standard $\Lambda$-CDM model probes reionisation history in a broader
range of cosmologies. Two extreme cases for $\tau$ are referred to in
the following as high-$\tau$ and low-$\tau$, corresponding to a MCMC
run using both temperature and polarisation at high and low multipoles
leaving the amplitude of lensing free and a run with temperature only
at high multipoles and temperature and polarisation at low multipoles
with curvature free, respectively\footnote{MCMC runs named
  $base\_omegak\_plikHM\_TT\_lowTEB$ and
  $base\_Alens\_plikHM\_TTTEEE\_lowEB$ are available on Planck Legacy
  Archive: \url{http://pla.esac.esa.int/pla/}}. The two corresponding
optical depth values are: $\tau \sim 0.11\pm 0.03$ and $\tau \sim
0.054\pm 0.021$.

We show in Fig.~\ref{fig:all} (upper panel) the current data and
constraints from direct observations of the ionised hydrogen fraction
$Q_\mathrm{HII}$ (symbols, from BOU15 and references therein), CMB (solid lines), and star-forming high-redshift
galaxies (dashed lines); namely, we plot the measurements from
line-of-sight observations as compiled in ROB15. We further add the
lower limit estimates of the neutral hydrogen fraction at $z \sim 7.7$
from two additional Ly-$\alpha$ emitters \citep{Faisst14} and the
neutral fraction estimate from the optical afterglow of the Swift
gamma-ray burst (GRB) 140515A at $z \sim 6.15$ \citep{Chornock14}.
The high-$\tau$ constraint from CMB (blue shaded area) is incompatible
with other reionisation constraints. We furthermore note that within
all the possible CMB models, those providing us with low optical
depths agree best with the data points. However, changes in $\tau$
only shift the best-fit CMB curves to high or low redshifts without
modifying their shapes. As a result, no satisfactory fit seems attainable
with the standard reionisation history used in CMB analyses.
Figure~\ref{fig:all} (middle panel) displays the evolution of the
SFR density $\rho_\mathrm{SFR}$ compiled by
{ROB15}. Overplotted are the best models from \cite{Rob15} (dashed
line) and the $\rho_\mathrm{SFR}$ derived from the best-fit CMB
models\footnote{To convert from ${\dot N}_{ion}$ to $\rho_\mathrm{SFR}$, we
  assume $\mathrm{log_{10}}(\langle f_{esc}\xi_{ion}\rangle) = 52.44$ as in
  ROB15.} with increasing values of $\tau$ from left (solid green
curve) to right (solid light-blue curve).  The dashed grey line
shows the regime where the ionisation fraction is set to unity, which is when ionisation is complete.  Finally, in Fig.~\ref{fig:all}
(lower panel) we plot the Thomson optical
depth derived from the ionising background due to star-forming
high-redshift galaxies as a function of redshift (dashed lines). We also display the optical
depth from the best-fit CMB models (solid lines), together with the
one-sigma measurement from PCP15 (shaded areas). As noted in {BOU15} and
{ROB15}, star-forming high-redshift galaxies provide enough ionising
background to obtain a Thomson optical depth of about 0.06, compatible
(on the lower side) with Planck standard $\Lambda$CDM constraints
($\tau \sim 0.066\pm 0.013$). Here again, the high-$\tau$ value (blue)
is definitively at odds with other astrophysical constraints.


\section{New parameterization of the reionisation history}

Current astrophysical constraints seem to favour low values of the
optical depth, regardless of the probe, with a sharp reionisation
transition between redshifts $\sim 7$ and $\sim
6$. Reionisation histories derived from the abundance of star-forming
galaxies suggest an asymmetric behaviour of $Q_\mathrm{HII}$ with
redshift. On the numerical side, recent simulations reproducing large
volumes of the Universe allow for multiple populations of ionising
sources, including early Population III stars and self-regulated UV
emitting sources \citep[e.g. ][]{ahn12,park13}.  These simulations
show that in the most general cases, the reionisation history
described in terms of { $x_e(z)$} shows a $z$-asymmetric behaviour
\citep[see e.g.  Fig.~3 in][]{park13}.

So far, CMB analyses in turn have assumed a redshift-symmetric
instantaneous reionisation history \citep[implemented e.g. in the
  publicly available code \texttt{CAMB},][]{Lewis99}. This widely used
reionisation parameterization is a tanh-based {expression}
$Q_\mathrm{HII}(z)\propto
\big(1+\tanh\big(\frac{(1+z)^3-(1+z_\mathrm{re})^3}{dz}\big)\big)/2$. The
two key parameters, $z_\mathrm{re}$ and $dz$, measure the redshift at
which the { $x_e$} reaches half its maximum (typically $1.08$,
not including the second reionisation of helium that occurs at {$z\sim
 3.5$)} and the
duration of reionisation, respectively. In this parameterisation
$z_\mathrm{re}$ is related one-to-one to $\tau$ for a fixed short
duration. As shown in the upper panel of Fig. \ref{fig:all}, the shape of the
tanh-based {expression} does not provide a good fit to the data. The middle panel of Fig.
\ref{fig:all} shows that the $\rho_\mathrm{SFR}$
evolution derived from CMB using a than-based parameterization does
not agree with the data, or barely agrees on a rather narrow range of
redshifts.

\begin{figure}[!h]
        \centering
        \includegraphics[width=8cm]{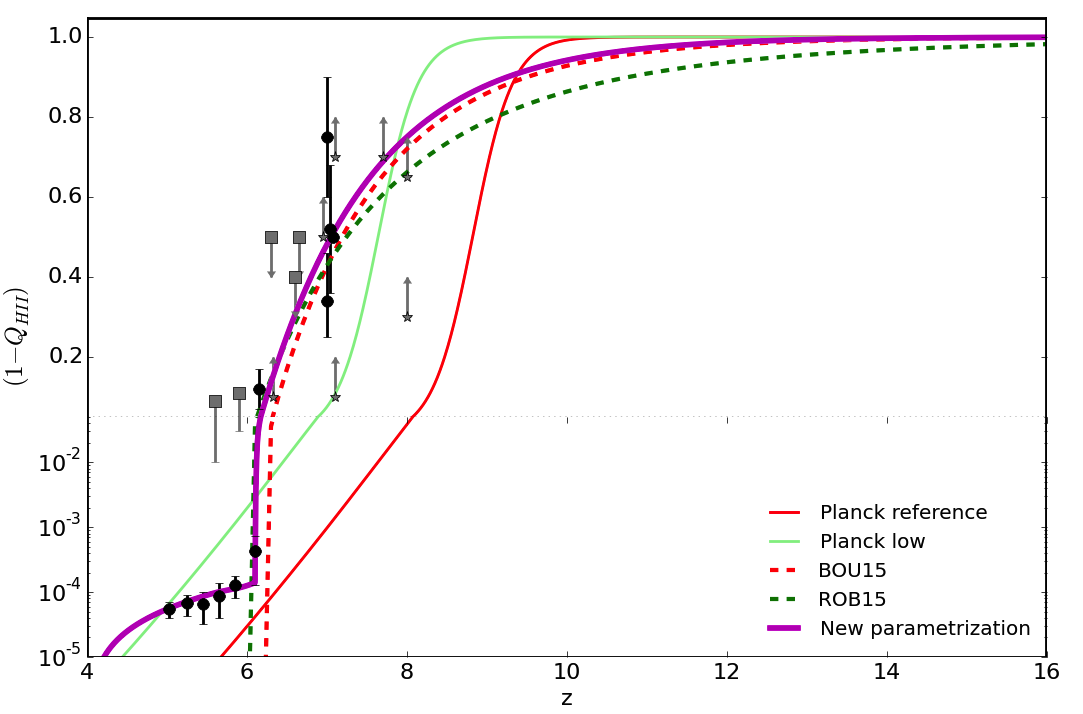}\\
\vspace{-0.1cm} \includegraphics[width=8cm]{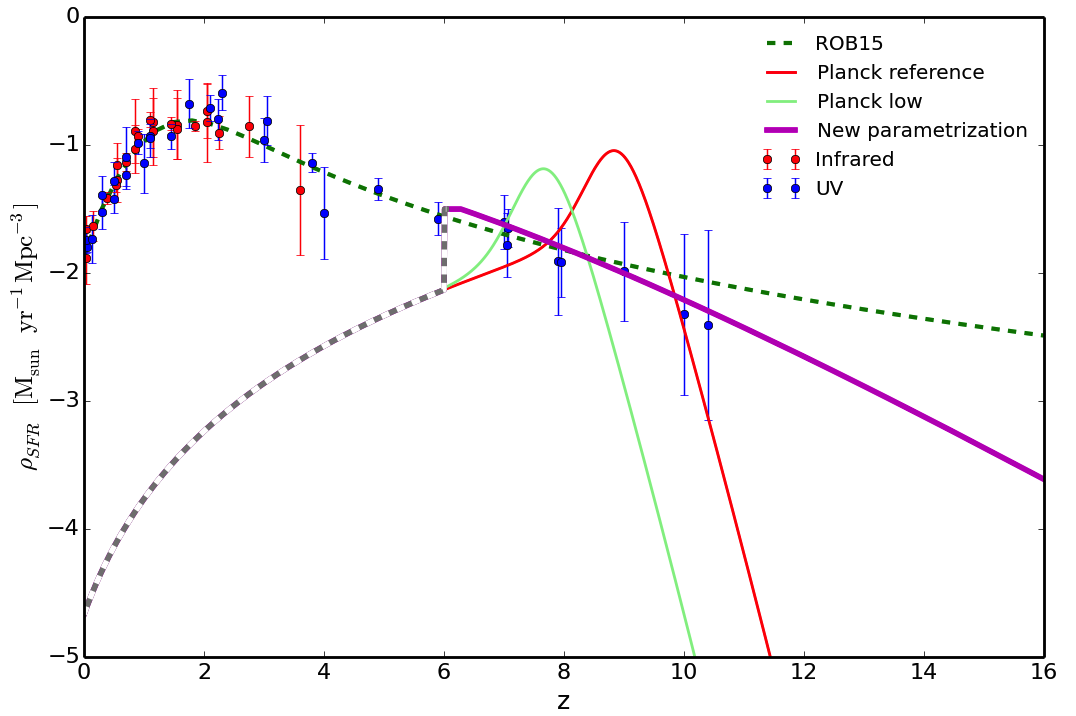}
\caption{Upper panel: Reionisation histories compared to low-redshift
  probes. Same as Fig. \ref{fig:all} with the magenta curve displaying
  the best-fit from the parameterization. Lower panel: Same as
  Fig. \ref{fig:all} with the magenta curve showing the associated
  derived $\rho_\mathrm{SFR}$.}
        \label{fig:approx}
\end{figure}

{The current astrophysical constraints on reionisation {suggest} that
tanh-based { histories} have difficulty reproducing the data even
in the low-$\tau$ regime.
We propose a new parameterisation that allows} for a first period
where reionisation can be slow and progressive, which is attributable to the
softer ionising photons produced by the first stars and primordial
dwarf galaxies, and a second period that can be faster leading, by $z
\sim 6$, to the completion of hydrogen and first helium reionisation
by quasars, for example, with harder ionising photons.  {This}
parameterisation is {motivated by} the behaviour of
$Q_\mathrm{HII}$ {as deduced from simulations and observations. At
  the end of reionisation, \cite{Fan06} show that the Gunn-Peterson
  optical depth evolves as $\tau_\mathrm{GP}\propto (1+z)^{\sim
    4.3}$. The neutral hydrogen fraction in turn evolves as
  $\tau_\mathrm{GP}/(1+z)^{3/2}$, so we consider $1-Q_\mathrm{HII}
  \propto (1+z)^{3}$ at low redshifts below 6. \cite{Fan06} also note
  that at redshifts 6 and above, $\tau_\mathrm{GP}$ grows much more quickly,
  possibly $\propto (1+z)^{>11}$. Such an accelerated phase is
  commonly observed in numerical simulations
  \citep{park13,Fern14}. We assume that this phase is in fact the
  end of an exponential phase. }

The parameterisation we {propose} here thus follows these two
asymptotic behaviours: a polynomial of order 3 {below} some pivot
redshift { ($z_p$)} and an exponential one {above}:
\begin{eqnarray}
\mathrm{if}\;\; z<z_p\; & 1-Q_\mathrm{HII}(z) &\propto (1+z)^{3}\\
\mathrm{if}\;\; z \geq z_p\; &   Q_\mathrm{HII}(z) &\propto exp(-\lambda (1+z))
.\end{eqnarray}
{The parameterisation is described by three free parameters: the pivot redshift $z_p$, the $Q_\mathrm{HII}$-value at the
  pivot, and the evolution rate in the exponential. Proportionality
  coefficients can be derived from these three parameters.} Depending
on the redshift and the $Q_\mathrm{HII}$-value at the pivot, the {  expression} is able to reproduce a large family of {reionisation
  histories}.

A forthcoming paper (Ili\'c et al., in prep.)  describes the parameterization and its effects on different
observations in more
detail, in particular the CMB.  We show in the
upper panel  of Fig.~\ref{fig:approx} the same data points and models as in
Fig.~\ref{fig:all}. We overplot our
parameterisation of the reionisation history fitted to the measured
values, i.e. excluding lower and upper limits. {The corresponding
  parameter values are: $z_p=6.1,\; Q_{HII}(z_p)=0.99986,\;
  \lambda=0.73$}. Figure~\ref{fig:approx} (upper panel) shows that the
new parameterisation performs as well as or slightly better than
constraints from ROB15 and BOU15 (dashed). It outperforms the
tanh-based reionisation expressions used in CMB analyses (solid lines)
while having the same value of $\tau$ as the Planck low model {and
  thus quasi-identical EE angular power spectra.  However, the redshift of reionisation as defined as $Q_{HII}(z_{reio})=0.5$
  is different in the two approaches: $z_{reio}=7.05$ for our parameterisation and
  $z_{reio}=7.65$  for the tanh expression.} Furthermore, the parameterisation reproduces the results from
BOU15 in an
economical way,  with only three free parameters. Using this best fit, we derived the
SFR density $\rho_\mathrm{SFR}$  from the parameterisation and compared it to
the results and compilation from ROB15 and to equivalent CMB-derived
values using the  tanh {expression} (Fig.~\ref{fig:approx} lower
panel). {In contrast to the latter}, the new parameterisation agrees
perfectly with the data points and with the ROB15 results in the
redshift range probed by star-forming high-redshift galaxies: {It
  reproduces a slow beginning of reionisation that accelerates to
  reach an ionisation fraction of about 1, around the pivot point
   and a slowly varying ionisation state until the
  present.}


\section{Conclusion}

Recent CMB observations favour lower values of the optical depth, in
agreement with values derived from ionising background estimates from
star-forming galaxies and low-redshift line-of-sight probes. While
this value is weakly sensitive to the modelling of the reionisation
history in CMB studies, the latter has so far been assumed to be a simple
step transition between neutral and fully reionised hydrogen. This
naive {parameterisation does not render} the behaviour of
$Q_\mathrm{HII}$ that is observed or derived from other probes. We propose
here a new {parameterisation motivated by simulations and
  observations of the reionisation fraction}. This new
parameterisation {reproduces a large family of reionisation
  studies in an economical way for CMB studies and furthermore} yields
a good fit to the line-of-sight measurements of the ionised fraction
and of the SFR densities. {Moreover, it provides a
  more realistic value of the redshift of reionisation.} We thus call
for implementing this parameterisation in future CMB analyses for
reionisation. Systematic studies of our parameterisation, in particular
in CMB studies, will be presented in a forthcoming paper.


\begin{acknowledgements}
  The authors thank B.E. Robertson for kindly providing us with his
  compilation of star formation rate densities. SI acknowledges
  support from the OCEVU Labex (ANR-11-LABX-0060) and the A*MIDEX
  project (ANR-11-IDEX-0001-02) funded by the ``Investissements
  d'Avenir'' French government programme managed by the ANR from LABEX
  OCEVU. The authors thank the referee for useful comments.
\end{acknowledgements}

\bibliographystyle{aa}

\bibliography{reio}

\end{document}